\documentclass{article}
\usepackage{spconf,amsmath,graphicx,hyperref}
\usepackage[T1]{fontenc}
\usepackage[utf8]{inputenc}
\usepackage{booktabs}
\usepackage[table]{xcolor}
\usepackage{colortbl}
\usepackage{amsmath,cite,url,amssymb}
\usepackage{graphicx}
\usepackage{color}
\usepackage{comment}
\usepackage{float}
\usepackage{array}
\usepackage{xurl}

\newcounter{row}
\newcommand{\nextrow}{\refstepcounter{row}\arabic{row}}

\title{Investigating the performance and energy costs of replicating \\Band-Split RNN for Music Source Separation}

\twoauthors
  {Paul Magron, Romain Serizel}
	{Université de Lorraine, CNRS\\
	Inria, LORIA, F-54000 Nancy, France}
  {Constance Douwes}
	{Centrale Med, Aix Marseille Univ\\
	CNRS, LIS, Marseille, France}

\begin{document}
\ninept
\maketitle
\begin{abstract}
  Band-split recurrent neural network (BSRNN) is a popular music source separation model that yields close to state-of-the-art results using reasonable computational resources and public datasets. It is therefore interesting from a reproducible research perspective, but achieving its performance is not straightforward since its full code is not available. In this paper, we conduct a replication of BSRNN via implementing the full pipeline. We extend the original paper's analysis by experimentally studying various design choices about data preprocessing, the optimization protocol, and architectural parameters. We report and discuss this project's energy cost, and we underline how its footprint could have been substantial lower upon availability of the full pipeline, which advocates for more reproducible research practices. To comply with this objective, we publicly release our code and pre-trained models.
\end{abstract}
\begin{keywords}
  Music source separation, band-split RNN, replication study, reproducible research, energy cost.
\end{keywords}

\section{Introduction}
\label{sec:intro}

Music source separation (MSS)~\cite{Cano2019} aims to extract the instrumental tracks that add up to form a music song.
MSS has experienced remarkable performance improvements over the past decade, with a trend towards deeper networks and intricate training protocols.



Despite this progress,  most recent methods (\textit{cf}. Table~\ref{tab:test_results}) suffer from several drawbacks from a \textit{reproducible research} perspective. For instance, winners of the latest music demixing challenge~\cite{Fabbro2024mdx23} such as Hybrid Demucs~\cite{Defossez2021hybrid} are bags of models, which can be tedious to train as they depend on multiple base architectures. The more advanced HT Demucs~\cite{Rouard2023htdemucs} outperforms Hybrid Demucs only when fine-tuned on a private dataset, which makes replication plain impossible. Systems such as BS-RoFormer require such large computing resources~\cite[Sec.4.4]{Lu2024} that retraining is prohibitively long when only a fraction of this capacity is available. These factors exacerbate the reproducibility crisis that plagues many fields of machine learning research, including music information retrieval~\cite{Six2018,McFee2019}.

\begin{table}[t]
  \center
  \scriptsize
  \caption{Separation performance (cSDR in dB, see Section~\ref{sec:protocol-eval}) for state-of-the-art MSS models trained and tested on MUSDB18-HQ.}
  \label{tab:test_results}
  \begin{tabular}
    {
      lcccc
      >{\columncolor{gray!10}}c
    }
    \toprule
                                            & \textbf{Vocals} & \textbf{Bass} & \textbf{Drums} & \textbf{Other} & \textbf{Average} \\
    KUIELab-MDX-Net~\cite{Kim2021kuielab}   & 8.97            & 7.83          & 7.20           & 5.90           & 7.47             \\
    Hybrid Demucs~\cite{Defossez2021hybrid} & 8.35            & 8.43          & 8.12           & 5.65           & 7.64             \\
    HT Demucs~\cite{Rouard2023htdemucs}     & 7.93            & 8.48          & 7.94           & 5.72           & 7.52             \\
    TFC-TDF UNet v3~\cite{Kim2023tfctdf}    & 9.59            & 8.45          & 8.44           & 6.86           & 8.34             \\
    BSRNN~\cite{Luo2023bsrnn}               & 10.01           & 7.22          & 9.01           & 6.70           & 8.24             \\
    SIMO-BSRNN~\cite{Luo2024simo}           & 9.73            & 7.80          & 10.06          & 6.56           & 8.54             \\
    BS-RoFormer~\cite{Lu2024}               & 10.66           & 11.31         & 9.49           & 7.73           & 9.80             \\
    DTTNet~\cite{Chen2024}                  & 10.12           & 7.45          & 7.74           & 6.92           & 8.06             \\
    SCNet~\cite{Tong2024scnet}              & 9.89            & 8.82          & 10.51          & 6.76           & 9.00             \\
    SCNet-large~\cite{Tong2024scnet}        & 10.86           & 9.49          & 10.98          & 7.44           & 9.69             \\
    \midrule
    BSRNN - ours                            & 8.91            & 7.46          & 8.07           & 5.22           & 7.42             \\
    BSRNN - alternative                     & 9.39            & 8.04          & 8.32           & 5.73           & 7.87             \\
    \bottomrule
  \end{tabular}
  \vspace{-2em}
\end{table}

Among state-of-the-art MSS methods, band-split recurrent neural network (BSRNN)~\cite{Luo2023bsrnn} is promising in this regard. Indeed, even though other recent methods such as DTTNet~\cite{Kim2023tfctdf} and SCNet~\cite{Tong2024scnet} exist, with available code and competing performance, BSRNN has fuelled future methodological developments~\cite{Chen2024,Luo2024simo,Lu2024}, and it was exploited for alternative tasks such as cinematic separation~\cite{Watcharasupat2023} or speech enhancement~\cite{Yu2023sebsrnn}. This large impact and popularity in the MSS community makes it a suitable candidate for a replication study. Unfortunately, while the BSRNN authors have shared some code for the model definition, there is no available official implementation of the \emph{full pipeline} that allows one to reproduce the results reported in the paper. This includes the code for data preprocessing, detailed training scripts with optimizer parameters, and evaluation functions. Yet, these are of paramount importance for reproducing any deep learning-based system's performance. Unofficial implementations exist and contain useful resources,\footnote{For instance, see {\scriptsize \url{github.com/amanteur/BandSplitRNN-PyTorch}},\\
{\scriptsize \url{github.com/crlandsc/Music-Demixing-with-Band-Split-RNN}},\\
or {\scriptsize \url{github.com/sungwon23/BSRNN}}.} but they either report substantially lower performance than that of the original, or adapt the pipeline to another dataset or task.

This motivates us to conduct a replication of BSRNN, which is in line with a growing interest within the machine learning community to value such studies: a noticeable example is the Reproducibility Challenge ({\scriptsize \url{reproml.org}}), which has become an official NeurIPS track in 2026.
In this paper, we propose to implement the full BSRNN pipeline as closely as possible to the original paper.
Our contributions are three-fold. First, we conduct extensive experiments to study various design choices, including data preprocessing, optimization protocol, and model parameters, which extends the original paper's analysis.
Second, drawing on related work in the field~\cite{Holzapfel2024greenmir,Douwes2023}, we report and discuss this study's energy cost. Indeed, deep learning-based research has a considerable environmental footprint and should move towards its systematic monitoring and reduction~\cite{Strubell2020}. We emphasize that MSS research could be more energy-effective upon favoring reproducibility.
Third, to comply with this objective, we publicly release our code and pre-trained models.\footnote{\scriptsize {\url{github.com/magronp/bsrnn}}}

\section{Model and training}
\label{sec:model}

This section describes the main building blocks of the BSRNN pipeline. For clarity, we focus on aspects or variants that need special care and clarification compared to the original paper~\cite{Luo2023bsrnn}, while practical experimental protocol details are described in Section~\ref{sec:protocol}.

\subsection{Original BSRNN model}
\label{sec:model-orig}

BSRNN takes as input the complex-valued short-time Fourier transform (STFT) of the mixture $\mathbf{X} \in \mathbb{C}^{F \times T}$, where $F$ and $T$ denote the number of frequency bands and time frames, respectively, and predict the STFT of a target source $\mathbf{S} \in \mathbb{C}^{F \times T}$. In practice, complex-valued STFTs are represented as stacks of their real and imaginary parts, since such a real-valued tensor can easily be processed with a neural network. BSRNN consists of the three following modules.

First, the \textit{band split} module decomposes the input STFT into frequency subbands with variable bandwidth, which are subsequently projected into a deep latent space with dimension~$N$. The band split scheme is designed and optimized for each source specifically, which yields instrument-specific models of different sizes.

Then, the \textit{sequence and band modeling} module applies two residual networks (one after the other), whose basic structure consists of a group normalization, a bidirectional long short-term memory, and a dense layer. These RNNs operate across time frames and bands, respectively, which optimally exploits the structure of audio/music signals by capturing dependencies across these dimensions. Such RNNs are stacked $R$ times to form a deeper network.

Finally, the \textit{mask estimation} module splits the output of the previous module into subbands, and each subband feature is fed to a multi-layer perceptron (MLP) that predicts a subband mask. The MLPs use a hidden size~$\mu \times N$ with $\mu=4$. Subband masks are then assembled into a fullband mask ${\mathbf{M} \in \mathbb{C}^{F \times T}}$ that is multiplied with the input STFT to yield the source estimate: $\hat{\mathbf{S}} = \mathbf{M} \odot \mathbf{X}$, which is finally reverted to time-domain via inverse STFT (iSTFT).

Many variants that build upon this base architecture have been proposed~\cite{Chen2024,Luo2024simo,Lu2024}. We refer the interested reader to our code for implementation details and analysis of the experimental results.

\subsection{Optimization}
\label{sec:model-optim}

BSRNN is trained using the following combination loss that consists of a time-domain and an STFT-domain term:
\begin{equation}
  \mathcal{L} = | \text{iSTFT}(\mathbf{S}) - \text{iSTFT}(\hat{\mathbf{S}}) |_1 + | \mathbf{S}_r - \hat{\mathbf{S}}_r |_1 +  | \mathbf{S}_i - \hat{\mathbf{S}}_i |_1,
  \label{eq:loss}
\end{equation}
where $|.|_1$ denotes the $\ell_1$ norm, and the subscripts $r$ and $i$ respectively denote the real and imaginary parts. In this paper we evaluate the impact of each term separately.

The original model is trained using a batch size of 2 and 8 GPUs in parallel, yielding a \textit{global} batch size $B=16$. Unfortunately, we do not have access to enough (large) GPUs to obtain the same $B$. As a result, we adjust the learning rate $\lambda$ in order to preserve the same \textit{effective} learning rate $\lambda / B$, ensuring similar gradient steps.

Training is conducted with a maximum of~$100$ epochs in the original paper, which we did not observe to be enough for convergence in most cases, thus we set this number at~$200$.
Besides, the authors apply early stopping ``when the best validation is not found in 10 consecutive epochs''~\cite[IV-A]{Luo2023bsrnn}, which does not clearly state which criterion (min loss vs. max evaluation metric - see Section~\ref{sec:protocol-eval}) is used. By default, we monitor the evaluation metric.

\subsection{Processing training and evaluation data}
\label{sec:model-data}

The original paper first preprocess training tracks with a source activity detector (SAD) that removes the silent regions. Then, training samples are generated on-the-fly by (i) extracting random 3~s-long chunks for each source, where sources are randomly mixed from different songs;\footnote{Even though this process yields inconsistent music mixtures - a phenomenon termed \emph{cacophony} - it was shown powerful for MSS~\cite{Jeon2024}.} (ii) randomly adjusting each chunk's energy in a $[-10, 10]$ dB range; and (iii) dropping each chunk with probability~0.1 to simulate silence sources. One epoch contains 20k samples.

We also consider an alternative strategy, without SAD preprocessing, and using augmentation techniques inspired from Open-Unmix~\cite{Stoter2019umx}, i.e., randomly swapping the channels, and rescaling the chunks' energy using a linear gain between~0.25 and 1.25.

At the evaluation stage (i.e., validation or test), songs are split into segments of 10~s with 10\% overlap, and the estimated chunks are assembled using a linear fader to smooth their edges~\cite{Defossez2021hybrid}. This technique is preferred over the original one~\cite[IV-C]{Luo2023bsrnn} since it yields similar results on average, but inference is faster by a factor of~6.

\section{Experimental protocol}
\label{sec:protocol}

\subsection{Dataset}
\label{sec:protocol-data}

As per the original paper, all experiments use the openly available MUSDB18-HQ dataset~\cite{Rafii2019musdb18hq}. It consists of 150 stereo songs sampled at 44100~Hz, with pairs of mixtures and corresponding four isolated sources: \texttt{vocals}, \texttt{bass}, \texttt{drums}, and \texttt{other}. It is split into 86, 14, and 50 songs for training, validation, and testing, respectively.

\subsection{Practical implementation}
\label{sec:protocol-implementation}

To accelerate prototyping, most experiments use a \textit{small} model, i.e., $N=64$ and $R=8$, while the original paper uses $N=128$ and $R=12$, which herein corresponds to the models denoted \textit{large}. We train models using the Adam algorithm with an initial learning rate $\lambda=10^{-3}$, decayed by~$0.98$ every $2$ epochs, and gradients are clipped by a maximum norm of 5. All other hyper-parameters are chosen as per the original paper, unless specified explicitly.

Small models are trained using 4 Nvidia RTX 2080 Ti (11~GB) GPUs, while large models use 2 Nvidia Tesla L40S (45~GB) GPUs. While using different GPU types affects energy comparison~\cite{Serizel2023} (see Section~\ref{sec:protocol-energy}), it allows us to optimally exploit our available hardware.

\subsection{Evaluation metrics}
\label{sec:protocol-eval}

We assess separation quality via two variants of the signal-to-distortion ratio (SDR)~\cite{Vincent2006}: the \textit{utterance} SDR (uSDR), computed by taking the mean SDR across whole songs~\cite{Fabbro2024mdx23}, and the \textit{chunk} SDR (cSDR), computed by taking the median SDR across 1 s-long chunks and across songs~\cite{Stoter2018}, using the museval toolbox~\cite{Stoter2021museval}.

\subsection{Monitoring energy consumption}
\label{sec:protocol-energy}

We track the energy during model training via the codecarbon toolbox~\cite{Courty2024codecarbon}. Since codecarbon tends to underestimate energy, we also follow the methodology of the green algorithms (GA) calculator~\cite{lannelongue2021green}, which approximates consumption based on hardware specifications.
This method was shown to yield results that are closer to actual power meter readings compared to codecarbon~\cite{jay2023experimental}. We consider a 3~W power per 8~GB of memory, and a power usage effectiveness factor of $1.5$, according to our computing platform's recommendation.
Lastly, we estimate the total GA project energy, including training additional model variants (not reported here due to space constraints), prototyping, and inference.

\section{Results}
\label{sec:results}

\subsection{Performance comparison}
\label{sec:results-perf}

We report the best model's validation uSDR across experiments in Table~\ref{tab:val}. Test results in terms of cSDR are reported in Table~\ref{tab:test_results}.

\begin{table*}[t]
  \center
  \caption{Model comparison in terms of best validation uSDR (in dB), number of parameters (in millions, ``-'' denotes the same value as in the preceding line), and estimated energy for training all sources (in kWh). Each line describes the difference with the base model at line~\ref{tabl:base}. Large deviations from the base average uSDR, as discussed in Section~\ref{sec:results-base}, are indicated with the $^\star$ symbol.}
  \label{tab:val}
  \begin{tabular}{
      cl|cccc
      >{\columncolor{gray!10}}c
      c|cc
    }

    \toprule

                                   &                                        & \multicolumn{5}{c}{\textbf{Validation uSDR}} & \textbf{Parameters} & \multicolumn{2}{c}{\textbf{Energy}}                                                                                      \\
                                   &                                        & \textbf{Vocals}                              & \textbf{Bass}       & \textbf{Drums}                      & \textbf{Other} & \textbf{Average}      &       & \textbf{codecarbon} & \textbf{GA} \\
    \nextrow\label{tabl:base}      & Base model: $N=64$, $R=8$              & $7.7$                                        & $6.1$               & $9.7$                               & $4.8$          & $7.1\phantom{\star}$  & 32.3  & 127                 & 168         \\
    \hline
    \nextrow\label{tabl:losst}     & Loss domain:  time                     & $7.9$                                        & $6.1$               & $9.4$                               & $4.9$          & $7.1\phantom{\star}$  & -     & 116                 & 153         \\
    \nextrow\label{tabl:losstf}    & Loss domain:  STFT                     & $7.9$                                        & $6.4$               & $9.6$                               & $4.9$          & $7.2\phantom{\star}$  & -     & 131                 & 173         \\
    \nextrow\label{tabl:acc}       & Accumulating gradients                 & $8.0$                                        & $5.8$               & $9.6$                               & $4.9$          & $7.1\phantom{\star}$  & -     & 129                 & 170         \\
    \nextrow\label{tabl:monit}     & Monitoring with the loss               & $7.5$                                        & $6.4$               & $9.3$                               & $4.8$          & $7.1\phantom{\star}$  & -     & 120                 & 159         \\
    \hline
    \nextrow\label{tabl:stft}      & STFT: window=4096, hop=1024            & $7.3$                                        & $5.9$               & $8.7$                               & $4.4$          & $6.6^\star$           & 37.1  & 58                  & 92          \\
    \nextrow\label{tabl:mask}      & Masker factor $\mu=2$                  & $7.9$                                        & $6.8$               & $9.4$                               & $4.4$          & $7.1 \phantom{\star}$ & 20.6  & 110                 & 151         \\
    \nextrow\label{tabl:mask1}     & Masker factor $\mu=1$                  & $7.5$                                        & $6.1$               & $9.8$                               & $4.7$          & $7.0\phantom{\star}$  & 16.8  & 114                 & 154         \\
    \hline
    \nextrow\label{tabl:silent}    & Silent target (instead of all sources) & $7.9$                                        & $6.6$               & $9.5$                               & $4.4$          & $7.1\phantom{\star}$  & 32.3  & 110                 & 146         \\
    \nextrow\label{tabl:nosad}     & No SAD, original augmentations         & $8.2$                                        & $6.6$               & $9.5$                               & $4.9$          & $7.3^\star$           & -     & 158                 & 210         \\
    \nextrow\label{tabl:nosad-alt} & No SAD, alt. augmentations             & $8.2$                                        & $6.9$               & $9.5$                               & $5.3$          & $7.5^\star$           & -     & 135                 & 179         \\
    \hline
    \nextrow\label{tabl:large}     & Large model: $N=128$, $R=12$           & $9.2$                                        & $7.3$               & $10.3$                              & $5.8$          & $8.2^\star$           & 146.7 & 230                 & 321         \\
    \nextrow\label{tabl:large30}   & + patience=30                          & $9.5$                                        & $7.8$               & $10.3$                              & $6.3$          & $8.4^\star$           & -     & 354                 & 495         \\
    \nextrow\label{tabl:largealt}  & + No SAD, alt. augmentations           & $9.6$                                        & $8.2$               & $10.0$                              & $6.4$          & $8.5^\star$           & -     & 326                 & 456         \\
    \bottomrule
  \end{tabular}
  \vspace{-1em}
\end{table*}

\vspace{-1em}
\subsubsection{Base model}
\label{sec:results-base}

\begin{figure}
  \centering
  \includegraphics[width=0.9\linewidth]{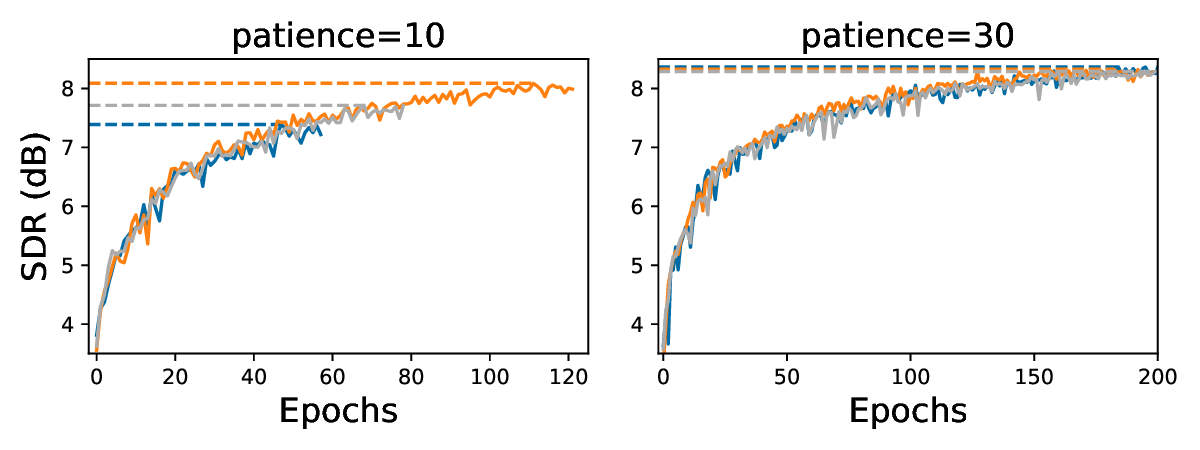}
  \vspace{-1.5em}
  \caption{Training the base model on the \texttt{vocals} track, with a patience of 10 (left) or 30 (right). Each color corresponds to a different run/seed, and the dashed lines correspond to each run's best uSDR.}
  \vspace{-1.5em}
  \label{fig:patience}
\end{figure}

As a preliminary experiment, we train a small model using three different random seeds, and display the validation uSDR over epochs in Figure~\ref{fig:patience} (left) for the \texttt{vocals} track (similar results are obtained for the other sources). We observe that all runs exhibit the same trend, but they yield some variability in terms of best uSDR, i.e., a standard deviation of about 0.3 dB for \texttt{vocals} (0.1 dB on average for all sources). These instabilities that stem from the different initial random seed ultimately cause training to stop at different epochs, which is an important reproducibility issue~\cite{Bouthillier2021}.

To alleviate it, we increase the patience parameter at~30, and we display the results in Figure~\ref{fig:patience} (right). We observe that the mean best uSDR is increased, but more importantly, its variance is largely reduced. Increasing patience is therefore effective to continue training and reduce the impact of the random seed onto the best uSDR. However, this comes at the cost of increasing the number of epochs, and consequently the energy, by a factor $2.3$.

Based on these findings, one could either reduce variance by tuning the random seed as any other hyperparameter~\cite{Bethard2022}, or average multiple runs / increase patience as done above. Nevertheless, to keep our experiments time and energy cost-effective and consistent with the original paper, we report a single run's results with a patience of~10, unless specified explicitly. Discrepancies inferior to the standard deviation observed above will be considered insignificant.

\subsubsection{Training parameters}

First, minimizing each term of the loss~\eqref{eq:loss} individually yields similar results to their combination (\textit{cf}. lines~\ref{tabl:base}-\ref{tabl:losstf}). This contrasts with previous studies that outlined the importance of time-domain training~\cite{Heitkaemper2020}. One explanation is that the STFT-domain term treats both the real \textit{and} imaginary parts separately, which implicitly enforces \textit{phase consistency}~\cite{LeRoux2008phase}, making it equivalent to a time-domain loss.

Instead of adjusting the learning rate as described in Section~\ref{sec:model-optim}, we can accumulate gradients over steps before performing descent, which artificially increases the global batch size. Results from line~\ref{tabl:acc} show that both strategies perform similarly. In what follows we adjust the learning rate as we observed it to be more stable when training larger models. Besides, while by default early stopping is performed via monitoring the uSDR (see Section~\ref{sec:model-optim}), we can instead monitor the validation loss. Both approaches yield similar results (\textit{cf}. line~\ref{tabl:monit}), but we chose the former as it allows training to continue for more epochs, which is beneficial in ensuring convergence.

\subsubsection{Original model parameters}

Originally, the STFT uses a 2048 sample-long Hanning window with a $75\%$ overlap ratio~\cite{Luo2023bsrnn}. An alternative setup using a larger window is common among MSS models~\cite{Stoter2019umx, Defossez2021hybrid}, but this yields poor result in our experiments (\textit{cf}. line~\ref{tabl:stft}). An explanation is that while a larger window increases the \textit{frequency resolution}, this has little impact on BSRNN since the band split scheme and projection occur early in the network (see Section~\ref{sec:model-orig}). Conversely, a larger hop size reduces the \textit{time resolution}, hence a performance loss that is more pronounced for the \texttt{drums} track, since percussive events are localized in time and thus require a refined time resolution to be properly modeled.

Besides, we reduce the original MLP masker factor $\mu$ from $4$ to $2$. While this negatively affects performance for the \texttt{drums} and \texttt{other} tracks, it substantially improves the \texttt{bass} results while also reducing this model's size, and it yields a similar average performance (\textit{cf}. line~\ref{tabl:mask}). Further decreasing $\mu$ at 1 degrades performance for the \texttt{bass}, but it cancels out the previous drop for the \texttt{drums} and \texttt{other} tracks (\textit{cf}. line~\ref{tabl:mask1}). This sheds light on a way to reduce model size, by adjusting such parameters for each instrument specifically.

\subsubsection{Data generation}
\label{sec:results-data}

Instead of randomly dropping each \textit{chunk} to simulate silent sources when generating data~\cite[IV-A-2]{Luo2023bsrnn}, we only drop the \textit{target} source, which improves performance for the \texttt{bass} track (\textit{cf}. line~\ref{tabl:silent}).
Altogether, removing the SAD preprocessing yields better performance for the \texttt{vocals} and \texttt{bass} tracks (line~\ref{tabl:nosad}). This suggests that there is room for improvement for our SAD implementation, since this preprocessing is alleged to greatly benefit the separation, although its impact is not evaluated specifically in the original publication~\cite{Luo2023bsrnn}.
Additionally applying the alternative augmentations further improves the \texttt{bass} and \texttt{other} performance, as observed at line~\ref{tabl:nosad-alt}.

\subsubsection{Large models}
\label{sec:results-large}

Using the original model size yields a large average uSDR improvement of 1.1~dB compared to the base model (line~\ref{tabl:large}). However, as displayed in Table~\ref{tab:test_results}, this implementation falls behind the original test results by 0.8~dB on average, which motivates further refining our pipeline. In an attempt to bridge this gap, and since we observed this large model was not fully converged (except for the \texttt{drums} track), we continue training after increasing the patience, which improves uSDR by 0.2~dB (line~\ref{tabl:large30}). We also retrain a large model using our alternative data generation process, which yields an additional improvement for the \texttt{bass} track (\textit{cf}. line~\ref{tabl:largealt}), consistently with previous experiments on small models.
As displayed in the last line of Table~\ref{tab:test_results}, this \textit{alternative} BSRNN model bridges half of the gap with the original test results compared to our implementation.

\subsection{Energy cost analysis}
\label{sec:results-energy}

\begin{figure}
  \centering
  \includegraphics[width=0.85\linewidth]{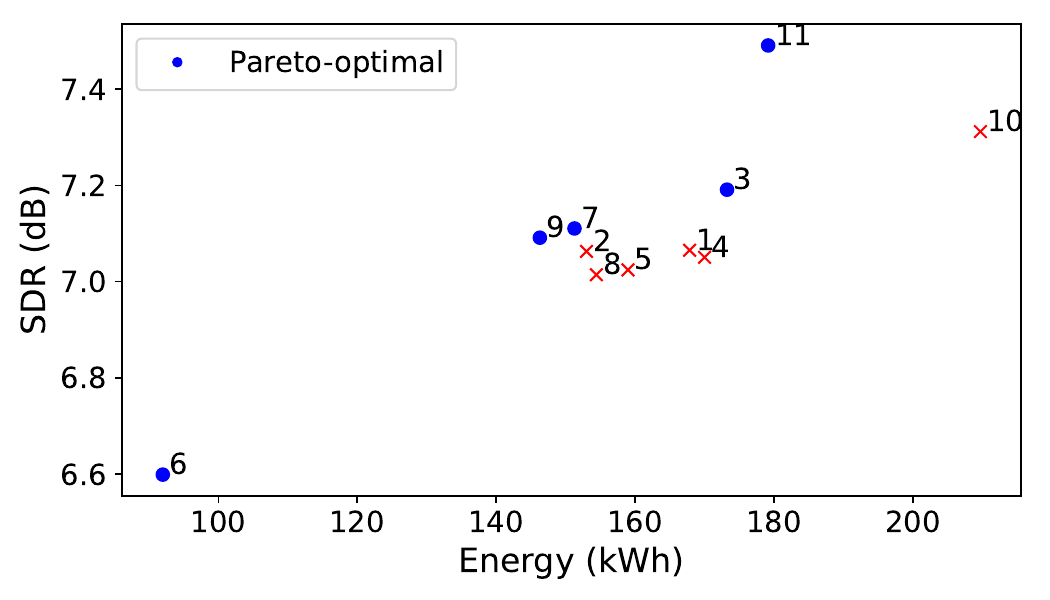}
  \vspace{-1.5em}
  \caption{Performance vs. training energy for small-size model variants, where each marker corresponds to the line number in Table~\ref{tab:val}.}
  \label{fig:pareto}
  \vspace{-1.7em}
\end{figure}

\subsubsection{Energy vs. performance}

From the last two columns of Table~\ref{tab:val},
we observe that both energy estimation techniques behave similarly in terms of relative variation between models. Consistently with previous studies~\cite{jay2023experimental}, the GA calculator yields larger values than codecarbon.

Most small-size model variants exhibit a similar energy cost, except for those that require either very few (line~\ref{tabl:stft}) or many epochs (lines~\ref{tabl:nosad}) to converge. Large models come with a substantial increased energy cost. Interestingly, the $0.2$ dB performance improvement of increasing the patience appears very limited when considering it costs an extra $54$\% energy (line~\ref{tabl:large} and~\ref{tabl:large30}). In contrast, training with alternative augmentations and no SAD reduces the energy toll, which is explained by a faster convergence of the \texttt{drums} model.

To further illustrate this, we display the validation uSDR against the GA energy in Figure~\ref{fig:pareto} (for readability we focus on small-size models). We notably highlight points that are optimal in a \textit{Pareto sense}~\cite{Douwes2023}, that is, such that there is no other point yielding both a higher SDR and a lower energy. This allows one to discard variants for which both a better performing and less costly alternative exists. For instance, one could favor training with the STFT-domain loss (\ref{tabl:losstf}) and/or using the alternative augmentation scheme (\ref{tabl:silent},~\ref{tabl:nosad-alt}), as well as reducing the masker size (\ref{tabl:mask}). Conversely, most alternative training strategies or the SAD preprocessing do not appear optimal in this sense. This highlights the importance of considering both performance \textit{and} energy consumption when comparing/selecting models.

\subsubsection{Discussion}

Training all the models discussed above amounts to $3$~MWh computed with GA, which represents about 7 times the energy cost of the single best model. While this ratio might seem reasonable at first glance, we also estimate the total project energy, as described in Section~\ref{sec:protocol-energy}. This amounts to 24.2~MWh, which is more than 53 times the energy consumption of training the best model, or 144 times that of the base model. To put things in perspective, this is equivalent to the yearly electricity consumption of about 15 persons in Europe.\footnote{\scriptsize \url{ec.europa.eu/eurostat/statistics-explained/index.php?title=Electricity_and_heat_statistics}}
Such a substantial toll advocates for systematic energy reporting and for more reproducibility. Indeed, this study's footprint would have been largely reduced upon availability of the code, documentation, and proper hyperparameter selection, as many trial-and-error experiments could have been avoided.

In all fairness, part of this energy cost is due to our own implementation errors, which resulted in, e.g., interrupted or redundant training runs. However, such pitfalls are common: for instance, a previous study~\cite{Strubell2020} reports that the electricity cost of the whole project is about $2000$ times that of training a single model, and the authors declare running almost $5000$ jobs in total, including many that crashed. In contrast, our study exhibits ratios at the lower end of the spectrum of similar projects, since it focuses on replicating a project rather than developing a new pipeline from scratch. Be that as it may, this total cost includes complementary experiments that we conducted in an attempt to bridge the performance gap, e.g., investigating alternative data augmentations, and further optimized models. We acknowledge that, strictly speaking, these do not correspond to replicating the original pipeline, therefore we cannot impute their cost to a lack of reproducibility - such a fine-grain allocation of the energy costs is unfortunately not possible through our computing platform. Nevertheless, most of these experiments would have been discarded upon easily reproducing the results.

Finally, even though these costs remain moderate compared to those of developing, e.g., large language models, encouraging sustainable practices remains important, especially since MSS research is also moving towards using such large language models~\cite{Wang2025unisep}.

\section{Conclusion}
\label{sec:conclu}

In this work, we have implemented a full BSRNN pipeline for music separation, extended the original paper's analysis, released an openly available model with competitive performance, and drew attention on the energy footprint of developing such a project.
Beyond this specific case study, our core contribution is to recall that reproducibility is a fundamental aspect of the scientific endeavour.
We encourage our colleagues to conduct more replication studies, and to adopt open research practices, notably via releasing their code with proper documentation~\cite{McFee2019}.
Through this work we hope to promote more transparent and sustainable practices.

\vfill\pagebreak
\section{Acknowledgements}
\label{sec:acknow}

All computation were carried out using the Grid5000 testbed, supported by a French scientific interest group hosted by Inria and including CNRS, RENATER and several Universities as well as other organizations. We thank J. Yu (author of BSRNN) for answering some of our implementation-related questions. We also thank C. Landschoot and S. Uhlich for fruitful discussion.

\bibliographystyle{IEEEtran}  
\bibliography{ref_icassp}

\end{document}